\documentclass[a4paper,showkeys,floatfix,aps,pre,longbibliography,superscriptaddress,reprint]{revtex4-2}
\usepackage{graphics,graphicx}
\usepackage{amsmath,amssymb}

\usepackage{graphics,graphicx}
\usepackage{dcolumn,bm}
\usepackage{psfrag}
\usepackage{xstring}
\usepackage{color}
\usepackage{comment}
\usepackage[colorlinks=true,breaklinks=true]{hyperref}

\renewcommand{\thesection}{\Roman{section}}

\renewcommand{\thesubsection}{\thesection~\Alph{subsection}}

\newcommand{\presidency}
{\affiliation{Department of Physics, Presidency University,  Kolkata-700073, India}}

\newcommand{\srm}
{\affiliation{Department of Physics, SRM University - AP, Amaravati,
 Andhra Pradesh - 522240, India}}

\newcommand{\srmcse}
{\affiliation{Department of Computer Science and Engineering, SRM University - AP, Amaravati,
 Andhra Pradesh - 522240, India}}

 \newcommand{\saha}
 {\affiliation{Condensed Matter Physics Division, Saha Institute of Nuclear Physics, Kolkata
700064, India}}

\begin{document}

\title{deGennes-Suzuki-Kubo Quantum
Ising Mean Field Dynamics: Applications
to Quantum Hysteresis, Heat Engines and Annealing}

\author{Soumyaditya Das}

\email{soumyaditya\_das@srmap.edu.in}
\srm

\author{Soumyajyoti Biswas}

\email{soumyajyoti.b@srmap.edu.in}
\srm
\srmcse

\author{Muktish Acharyya}

\email{muktish.physics@presiuniv.ac.in}
\presidency

\author{Bikas K. Chakrabarti}
\email{bikask.chakrabarti@saha.ac.in}
\saha

\begin{abstract} We briefly review the early development of the mean-field dynamics
for cooperatively interacting  quantum many-body systems, mapped to
pseudo-spin (Ising-like) systems. We start with (Anderson, 1958)
pseudo-spin mapping of the BCS (1957) Hamiltonian of
superconductivity, reducing it to a mean-field Hamiltonian of XY (or
effectively Ising) model in a transverse field. Then get the
mean-field estimate for the equilibrium gap in the ground state energy
 at different temperatures  (gap disappearing at the transition
temperature), which fits Landau’s (1949) phenomenological theory of
superfluidity. We then present in detail a general dynamical extension (for
non-equilibrium cases) of the mean-field theory of quantum Ising
systems (in a transverse
field), following de Gennes’ (1963) decomposition of the mean field
into orthogonal classical cooperative (longitudinal)  component and
the quantum (transverse) component, with each of the component
following Suzuki-Kubo (1968) mean-field dynamics.  Next we discuss its
applications to quantum hysteresis in Ising magnets (in presence of
oscillating transverse field), to quantum heat engines (employing
transverse Ising model as working fluid), and to the quantum annealing
of the Sherrington-Kirkpatrick (1975) spin glass by tuning down (to
zero) the transverse field which provided us a very fast computational
algorithm leading to ground state energy values converging to the best
known analytic estimate for the model. Finally, we summarize the main
results obtained and conclude about the effectiveness of the
deGennes-Suzuki-Kubo mean-field equations for the study of various
dynamical aspects of quantum condensed matter
systems.
\end{abstract}


\maketitle

\section{Introduction}

The first significant  theory of cooperatively interacting  many-body
quantum condensed matter systems had been the Bardeen–Cooper–Schrieffer
(BCS) theory \cite{bardeen} of superconductivity in 1957. The theory had been
spectacularly successful in explaining several experimentally observed
outstanding features of the superconducting phase of condensed matters
(without any external magnetic field). Next year, in an intriguing
publication \cite{anderson}, Anderson showed that BCS theory is in effect a  mean
field theory of the Cooper-pairwise (cooperatively) interacting electrons
(mapped to XY spins) in presence of a quantum (transverse) field of
strength given by the free electron energy (effectively to the mean field
theory of Ising systems in transverse field). The strength of the mean
field on each of the pseudo spins (corresponding to the wave-vectors  of
the electrons near the Fermi surface) would be given by the square-root of the sum of the squares of the free
electron energy and the sum of the square of the Copper pair energies (gap
parameter). This would lead to the self-consistent gap equation of
the BCS. Assuming that the gap vanishes at the transition temperature, one
can find the celebrated BCS relation between the gap magnitude at zero
temperature and  the superconducting transition temperature (see section II).

Though the BCS theory had been  a mean field one for a quantum cooperative
system (effectively a mean field theory of  XY or Ising system in
transverse field), the transition (at $T_c$, from a superconducting
phase to a normal phase) was not a quantum transition; it was driven by the
(classical) thermal fluctuations. Also it did not consider any dynamic
aspect of the transition, rather it is an equilibrium transition. Soon, a
mean field theory of quantum transitions (driven by quantum fluctuations)
for many-body condensed matter systems (e.g., in  the context of 
ferro-para electric transitions in  Potassium Dihydrogen Phosphate or KDP) was developed. Model systems of such transitions (driven by quantum tunneling of
hydrogen ions or protons in oxygen double-wells at each site) were
developed by de Gennes and others (see e.g., \cite{deGennes1963,Brout1966}). Here, the
`softening' of long wavelength or low frequency collective modes of
protons would give rise to dipole moments inducing dipole-dipole
interaction and ferroelectricity  in KDP. They calculated (see also \cite{eliot1})
the collective  tunneling frequency in such simplified models which can
induce softening of those collective modes (driven by the tunneling field
at different temperatures, including zero temperature for purely quantum
one) in such cooperative order-disorder trasition systems. This then led
generally to the study of static behavior of quantum phase transitions in many-body systems
employing the transverse field Ising models (see e.g., \cite{eliot2,pfu,stin,bkc}).
The mean field study of the transitions in these transverse field Ising
models had wider applications to tunneling mode softening studies in
quantum ANNNI model to study various  modulated phases in materials with
regularly competing interactions (see e.g., \cite{barbar} for the earliest study and \cite{SenChakrabarti1989}
for a mean field study of the softening of the corresponding dynamic modes).
First reports on similar  studies in transverse Ising models were made in
\cite{fisher} for random transverse field case and in \cite{sen} for spatially modulated
longitudinal field case. The earliest study of the static transition
properties of Mattis and Edwards-Anderson Ising spin glass in transverse
filed was reported in \cite{bkc81}.

Soon, dynamical behavior of transverse field Ising systems under
externally driven fields received the attention and dynamic hysteresis
in pure ising magnets, employing the Suzuki-Kubo mean field dynamics \cite{SuzukiKubo1968},
in both classical (see e.g., \cite{AcharyyaPhysicaA1994}) and quantum \cite{AcharyyaJPhysA1994} Ising systems (where
the longitudinal field and the tunneling or transverse field varied
periodically with time in the respective cases). Afterwards, this mean
field quantum dynamics for heat engines with pure Ising system as working fluid (with
both longitudinal and transverse fields varying cyclically as the bath
temperature changes from that of the heat source  to that of the sink) was studied
\cite{acharyya3} to show that quantum heat engine efficiencies can approach the Carnot
value while the corresponding classical engine can not. Recently, this
deGennes-Suzuki-Kubo mean field Ising dynamics has been applied to the
celebrated problem of quantum annealing \cite{X1,das_rmp,tamura,rajak,ency} to estimate the ground state
energy of the randomly frustrated spin glass model of the
Sherrington-Kirkpatrick (SK) variety \cite{SK1975,binder_young} where the best analytic estimate
(up to tenth decimal place exists \cite{Oppermann2007}) and the numerical study of the
quantum annealing gives \cite{das2} very fast convergence (with comparable
scaling results on the fluctuations) as  seen also  \cite{DasPRE2025} for the classical
annealing (just with Suzuki-Kubo dynamics).

We intend to develop here the most general form of deGennes-Suzuki-Kubo
quantum Ising mean field dynamical equations in section (III A)  and
then study them numerically for the special cases of dynamic hysteresis in
quantum Ising systems (in section III B), for the increased efficiencies of
the quantum Ising heat engines (in section III C) and for the quantum
annealing study the ground state energy for the SK model (in section
III D). Finally, we will summarize (in section IV) the main results discussed
here in different cases and highlight the elegance and prospect of the
deGennes-Suzuki-Kubo quantum Ising mean field dynamical equation to study
various many-body dynamical behavior of quantum condensed matter systems.

\section{Anderson’s Pseudo-Spin Mapping of
BCS Hamitonian \& Mean Field Theory
of BCS Superconductivity}

\subsection{Landau's Phenomenological Theory of Superfluidity and Superconductivity}

Helium-4 ($He^4$) when cooled below $4.2$K,  becomes liquid. If liquid $He^4$ was placed in a container, with the liquid filled almost to the edge, then there would be constant dripping of $He^4$ from the container. This indicates the disappearance of friction or viscosity of $He^4$ (in the surface layers). There were attempts to explain this from Bose-Einstein condensation. But this was not a system of free particles, but an interacting system. Therefore, it is not a Bose gas, but a Bose liquid. Hence, the superfluidity cannot be explained through Bose-Einstein condensation. Landau suggested the following scenario (see e.g., \cite{landau1,landau2}). 

For free particles, the dispersion relation reads $\epsilon_k \sim k^2$. However, such a free particle dispersion cannot show superfluidity. Landau assumed that for interacting Boson systems the dissipation relation may become $\epsilon_k \sim ck$ (for electron systems $\epsilon_k \sim \Delta+ck^2$ is possible) and that might lead to frictionless flow. He considered the superfluid system  consisting of the (quantum) fluid placed in contact with a (classical) massive wall or surface. If some friction is encountered in the flow, then energy of the fluid and hence the momentum has to change. This change in momentum is to be taken up by the classical mass in contact with the liquid. 

\begin{figure}[tbh]
\begin{center}
 \includegraphics[width=0.85\linewidth]{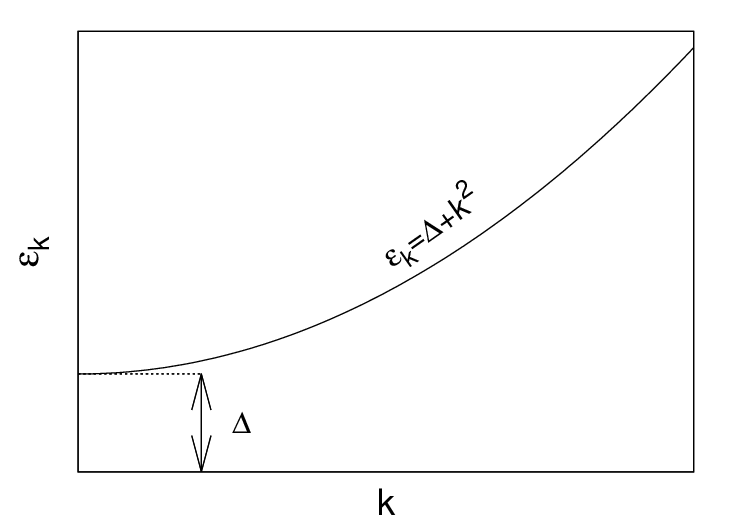}
\end{center}
   \caption{ 
  A schematic of the dispersion of low energy excitations in a quantum many-body system.}
\label{epsilon_k}
\end{figure}
If it can be shown that this heavy mass cannot accept every excitation satisfying the energy and momentum conservations, then the fluid has no other choice but to flow without friction. 

If $v$ and $v^{\prime}$ denote velocities of the classical
 mass before and after energy exchange, then from energy conservation
\begin{eqnarray}
\frac{1}{2}Mv^{\prime 2}=\frac{1}{2}Mv^2+\epsilon_k,
\label{app3.44} 
\end{eqnarray}
and for momentum conservation
\begin{eqnarray}
Mv^{\prime}=Mv+\hbar k .\nonumber
\end{eqnarray}
Squaring and dividing  the above equation by $2M$ we get
\begin{eqnarray}
\frac{1}{2M}M^2v^{\prime 2}=\frac{1}{2}Mv^2+\frac{1}{2M}\hbar^2k^2+\hbar vk .\nonumber
\end{eqnarray}
Subtracting this from eqn.(\ref{app3.44}), we get
\begin{eqnarray}
\epsilon_k=\frac{1}{2M}\hbar^2k^2+\hbar vk. \nonumber
\end{eqnarray}
Since $M$ is very large, the above energy-momentum conservation is satisfied for
\begin{eqnarray}
\left.\frac{\epsilon_k}{k}\right\arrowvert_{min}=\hbar v.
\label{app3.45}
\end{eqnarray}
If $\epsilon=ck^2$, then the above condition is satisfied for all non-vanishing velocities, indicating the possibility of exchange of energy and  momentum. Hence no  super-fluidity. But if $\epsilon_k=ck$ (as in the cases of phonon and roton excitations) then only for  $v>c$, the condition is satisfied. For lower values of $v$, one gets  flow without resistance or superfluidity. The other case of interest is $\epsilon_k=\Delta+k^2$ where one also gets superfluidity for $\Delta \ne 0$ and this can occur for the BCS theory of superconductivity discussed next.

\subsection{Mean Field Theory of BCS Superconductivity}

The resistivity in some materials  practically vanishes  when the temperature is reduced below a critical value. These materials then also become completely diamagnetic. This is called superconductivity. Superconductivity can not be a property of free electrons. In a lattice, electrons are of course not free. Apart from the Coulomb repulsion among them they face a fluctuating periodic potential (electron-phonon scattering). It is  observed that  replacement of atoms by their isotopes induce changes in the superconductivity onset temperature. This is called the isotope effect, which often relates the critical temperature ($T_c$) with the mass ($M$) of the atoms by a relation like $M^{1/2}T_c=constant$.  When we change the atomic mass, we basically do nothing to the electron but only change the phonon modes. This led  to the belief that superconductivity is  due to electron-electron interactions, mediated by phonons.  Cooper showed that the maximum value of the effective  interaction between the electron pairs can become attractive when the electron pairs are near the Fermi surface and they have opposite spin states and momentum vectors. The BCS Hamiltonian can be written, with $k$, $k^{\prime}$ denoting the momenta of electrons and $\tilde{k}$ for phonons and $C^{\dagger}$, $C$ denoting respectively the electron creation and destruction operators, as (see for e.g., \cite{anderson,kittel})

\begin{eqnarray}
H=\sum\limits_{k}\epsilon_k^0C_{k}^{\dagger}C_{k}+\sum\limits_{k,k^{\prime},\tilde k}V_{kk^{\prime}\tilde k}C_{k+\tilde k}^{\dagger}C_{k^{\prime}-\tilde k}^{\dagger}C_kC_{k^{\prime}}.
\label{app4.2}
\end{eqnarray}
 We assume here $V_{kk^{\prime}\tilde k}\equiv -V$, the maximum value of the effective attractive interaction when $k$ and $k^{\prime}(=-k)$ pairs (with opposite spin states) are formed and $k$'s are near Fermi vector. Then the Hamiltonian can be written as
\begin{eqnarray}
H=\sum\limits_k\epsilon_k^0C_k^{\dagger}C_k-V\sum\limits_{kk^{\prime}}C_{k^{\prime}}^{\dagger}C_{-k^{\prime}}^{\dagger}C_kC_{-k}.
\label{app4.3}
\end{eqnarray}
Because of pairing of electrons, we have ($k,-k$) and ($k^{\prime},-k^{\prime}$) pairs, both having the same energy $\epsilon_k^0$. Hence the Hamiltonian takes the form
\begin{eqnarray}
H=\sum\limits_k\epsilon_k^0\left(C_k^{\dagger}C_k+C_{-k}^{\dagger}C_{-k}\right)-V\sum\limits_{kk^{\prime}}C_{k^{\prime}}^{\dagger}C_{-k^{\prime}}^{\dagger}C_kC_{-k}. \nonumber
\end{eqnarray}
We choose $\sum\limits_k\epsilon_k^0=0$. Therefore,
\begin{equation}
H=-\sum\limits_k\epsilon_k^0\left(1-C_k^{\dagger}C_k-C_{-k}^{\dagger}C_{-k}\right)-V\sum\limits_{kk^{\prime}}C_{k^{\prime}}^{\dagger}C_{-k^{\prime}}^{\dagger}C_kC_{-k}. \nonumber
\end{equation}
The last term is still not in a diagonal form. We intend to map this Hamiltonian (in the lowest energy states) to a pseudo spin  Hamiltonian.

Let us consider only the low lying states of this Hamiltonian, namely the electron pair occupied ($\left\arrowvert-\rangle\right.$) and pair unoccupied ($\left\arrowvert+\rangle\right.$):
\begin{eqnarray}
\left\arrowvert+\rangle\right.\equiv\left\arrowvert 0_k,0_{-k}\rangle\right. \quad \left\arrowvert-\rangle\right.\equiv\left\arrowvert 1_k,1_{-k}\rangle\right. \nonumber
\end{eqnarray}
Hence,
\begin{eqnarray}
\left(1-C_k^{\dagger}C_k-C_{-k}^{\dagger}C_{-k}\right)\left\arrowvert+\rangle\right.=\left\arrowvert+\rangle\right. \nonumber \\
\left(1-C_k^{\dagger}C_k-C_{-k}^{\dagger}C_{-k}\right)\left\arrowvert-\rangle\right.=-\left\arrowvert-\rangle\right.. \nonumber
\end{eqnarray}
We therefore make the correspondence $(1-C_k^{\dagger}C_k-C_{-k}^{\dagger}C_{-k}) \equiv \sigma_k^z$.

Since $C_k^{\dagger}C_{-k}^{\dagger}\left\arrowvert +\rangle\right.=\left\arrowvert-\rangle\right.$,$\quad C_k^{\dagger}C_{-k}^{\dagger}\left\arrowvert -\rangle\right.=0$, $\quad C_{-k}C_{k}\left\arrowvert -\rangle\right.=\left\arrowvert +\rangle\right.$
and $\quad C_{-k}C_{k}\left\arrowvert +\rangle\right.=0$,

we immediately identify its correspondence with raising and lowering operators $\sigma^+/\sigma^-$:
\[\sigma^{\pm}=\sigma^x\pm i\sigma^y=\left( \begin{array}{cc}
0 & 0  \\
2 & 0  \end{array} \right) \quad \mbox{or}\quad \left( \begin{array}{cc}
0 & 2  \\
0 & 0  \end{array} \right),\]
and therefore
\begin{eqnarray}
C_k^{\dagger}C_{-k}^{\dagger}=\frac{1}{2}\sigma_k^-, \quad C_{-k}C_k=\frac{1}{2}\sigma_k^{+}.
\end{eqnarray}

In terms of these spin operators we finally arrive at
\begin{eqnarray}
H=-\sum\limits_k\epsilon_k^0 \sigma_k^z-\frac{V}{4}\sum\limits_{kk^{\prime}}\left(\sigma_k^x \sigma_{k^{\prime}}^x+\sigma_k^y \sigma_{k^{\prime}}^y\right)
\label{app4.4}
\end{eqnarray}
or,
\begin{eqnarray}
H = -\sum\limits_k{\vec h_k}\cdot {\vec \sigma_k}, 
\label{app4.5}
\end{eqnarray}
where the effective field ${\vec h_k}$ components are:
\begin{eqnarray}
h_k^{x}=\frac{V}{4}\sum\limits_{k^{\prime}}\langle \sigma_{k^{\prime}}^x\rangle; \quad h_k^{y}=\frac{V}{4}\sum\limits_{k^{\prime}}\langle \sigma_{k^{\prime}}^y\rangle ;  \quad h_k^{z}=\epsilon_k^0.
\label{eqn8}
\end{eqnarray}
 As the $\langle\sigma_x\rangle$ and $\langle\sigma_y\rangle$ are symmetric we consider only one component (with proper counting) and hence $\left\arrowvert {\vec h_k}\right\arrowvert=\sqrt{{\epsilon_k^{0}}^2 +\frac{V^2}{4}\left(\sum\limits_{k^{\prime}}\langle \sigma^x_{k^{\prime}}\rangle\right)^2}$. The pseudo spin $\vec{\sigma_k}$ will therefore be in the direction of the field $\vec{h_k}$ and its magnitude will depend on the temperature.

\bigskip{}
\noindent{\bf{\underline{At $T= 0$}:}}
\bigskip{}

\noindent Here the spin magnitude will be its maximum and hence

\begin{eqnarray}
\langle {\vec \sigma_k}\rangle=\frac{{\vec h_k}}{\left\arrowvert {\vec h_k}\right\arrowvert} 
\label{app4.5a0}
\end{eqnarray}
Let
\begin{eqnarray}
\langle \sigma_k^x\rangle=\sin \theta_k=\frac{\langle h_k^x\rangle}{\left\arrowvert {\vec h_k}\right\arrowvert} \quad \mbox{and} \quad \langle \sigma_k^z\rangle=\cos \theta_k=\frac{\langle h_k^z\rangle}{\left\arrowvert {\vec h_k}\right\arrowvert}, \nonumber
\end{eqnarray}
Hence
\begin{widetext}
\begin{eqnarray}
\langle \sigma^x_k\rangle=\frac{\sum\limits_{k^{\prime}}\frac{V}{2}\langle \sigma_{k^{\prime}}^x\rangle}{\sqrt{{\epsilon_k^0}^2+\frac{V^2}{4}\left(\sum\limits_{k^{\prime}}\langle \sigma_{k^{\prime}}^x\rangle\right)^2}}\quad \mbox{and} \quad \langle \sigma_k^z\rangle=\frac{\epsilon_k^0}{\sqrt{{\epsilon_k^0}^2+\frac{V^2}{4}\left(\sum\limits_{k^{\prime}}\langle \sigma_{k^{\prime}}^x\rangle\right)^2}} \nonumber
\end{eqnarray}
\end{widetext}
If we define

\begin{eqnarray}
\Delta=\frac{V}{2}\sum\limits_{k^{\prime}}\langle \sigma_{k^{\prime}}^x\rangle, 
\label{app4.5a}
\end{eqnarray}
then we can write
\begin{widetext}
\begin{eqnarray}
\langle \sigma_k^x\rangle=\sin \theta_k=\frac{\Delta}{\sqrt{{\epsilon_k^0}^2+\Delta^2}} \quad \mbox{and} \quad \langle \sigma_k^z\rangle=\cos \theta_k=\frac{\epsilon_k^0}{\sqrt{{\epsilon_k^0}^2+\Delta^2}} 
\label{app4.5b}
\end{eqnarray}
\end{widetext}
Putting eqn. (\ref{app4.5b}) on eqn. (\ref{app4.5a}) we get the self consistent equation for $\Delta$:
\begin{eqnarray}
\Delta=\frac{V}{2}\sum\limits_{k^{\prime}}\frac{\Delta}{\sqrt{{\epsilon_{k^{\prime}}^0}^2+\Delta^2}}.
\label{app4.6}
\end{eqnarray}
 We can express this gap in eqn. (\ref{app4.6}) as an integral in the following form
\begin{eqnarray}
\Delta=\frac{V}{2}\int\frac{\rho(\epsilon_F)\Delta d\epsilon}{\sqrt{{\epsilon}^2+\Delta^2}}, \nonumber
\end{eqnarray}
where $\rho(\epsilon_F)$ is the density of states at the Fermi level. Therefore, the so called {\it gap equation} takes the form
\begin{eqnarray}
\frac{V}{2}\int\frac{\rho(\epsilon_F) d\epsilon}{\sqrt{{\epsilon}^2+\Delta^2}}=1. 
\end{eqnarray}

\noindent The ground state of the Hamiltonian (\ref{app4.5}) occurs when ${\vec \sigma_k}$ is aligned in the direction of ${\vec h_k}$. The  excited state will be when the alignment of ${\vec \sigma_k}$ is opposite to ${\vec h_k}$. Hence the change in energy is $2\left\arrowvert {\vec h_k}\right\arrowvert$. Therefore,

\begin{equation}
\epsilon_k=2\left\arrowvert {\vec h_k}\right\arrowvert=2\sqrt{{\epsilon_k^0}^2+\Delta^2}.
\label{appnew}
\end{equation}

When $k \to 0$, $\epsilon_k^0\sim k^2\to 0$, but  $\epsilon_k \ne 0$. Hence $\Delta$ represents the zero temperature  energy gap, independent of $k$. As discussed in the previous section, the non-vanishing value of this gap $\Delta$ ensures a "superfluid" like flow of the (charged) electrons and hence superconductivity.

\bigskip{}
\noindent{\bf{\underline{At $T\ne 0$}:}}
\bigskip{}

\noindent Here, unlike in (\ref{app4.5a0}), the average spin magnitude will be given by $\tanh\left(\frac{\left\arrowvert {\vec h_k}\right\arrowvert}{k_BT}\right)$. Hence
\begin{eqnarray}
\langle {\vec \sigma_k}\rangle=\frac{{\vec h_k}}{\left\arrowvert {\vec h_k}\right\arrowvert}\tanh\left(\left\arrowvert {\vec h_k}\right\arrowvert/k_BT\right), \quad \left\arrowvert{\vec h_k}\right\arrowvert=\sqrt{{\epsilon_k^0}^2+\Delta^2}, \nonumber
\end{eqnarray}
and consequently the generalization of eqn. (\ref{app4.6}) would be
\begin{eqnarray}
\Delta=\frac{V}{2}\sum\limits_{k^{\prime}}\tanh\left(\frac{\left\arrowvert {\vec h_{k^{\prime}}}\right\arrowvert}{k_BT}\right)\sin\theta_{k^{\prime}}; \quad \sin\theta _k=\frac{\Delta}{\left\arrowvert {\vec h_k}\right\arrowvert}. 
\label{app4.6a}
\end{eqnarray}
The gap $\Delta$ is now a function of temperature $T$ and if we define the critical temperature $T_c$ as the temperature where the gap $\Delta$ vanishes (see previous section), then
\begin{eqnarray}
\frac{V}{2}\sum\limits_{k^{\prime}}\frac{1}{\epsilon_{k^{\prime}}^0}\tanh\frac{\epsilon_{k^{\prime}}^0}{k_BT_c}=1. 
\label{app4.6b}
\end{eqnarray}
Numerically solving equations (\ref{app4.6a}) and (\ref{app4.6b}), we get
\begin{eqnarray}
2\Delta\left(T=0\right)=3.5T_c.
\label{app4.9}
\end{eqnarray}

For most conductors (like Sn, Al, Pt), this relation is seen to be fairly
accurate.


\section{deGennes-Suzuki-Kubo Quantum Ising Mean Field Dynamical Equation}

In the previous section, we have seen that the BCS theory could be seen as a
mean field theory of a quantum XY system (described by Hamiltonian (\ref{app4.4})),
which in effect (because of the assumed symmetries, following eqn. (\ref{eqn8}),
for the thermal averages $\langle\sigma^x\rangle$ and $\langle\sigma^y\rangle$), becomes in effect a mean
field theory of an Ising model in transverse field. Also, in the BCS
theory, we did not consider any dynamics of that quantum Ising model.

In this section we therefore give a general dynamical mean field theory of
Ising systems in transverse field (given in the next section III A) and
then discuss special cases of mean field theories for quantum Ising
hysteresis in section III B, quantum Ising heat engine in section
III C, and quantum annealing of the Sherrington-Kirkpatrick Ising spin
glass model in section III D.

\subsection{Mean Field Dynamical Equation for a General Quantum Ising System}

The Hamiltonian of a general Ising system in the presence of time ($t$) dependent transverse ($\Gamma(t)$)
and longitudinal magnetic field ($h(t)$) reads as

\begin{equation}
 H(t) = - \sum_{i,j} J_{ij}\sigma_i^z\sigma_j^z
-h(t) \sum_i\sigma_i^z -\Gamma(t) \sum_i \sigma_i^x, 
\label{skd1}
\end{equation}

\noindent where $\vec {\sigma}$ denotes the Pauli
spin vector, $h(t)$ and $\Gamma(t)$ are the time dependent external
longitudinal and transverse field, respectively.
$J_{ij}$ denotes the strength of the interaction that can generally be random with specific distributions $P(J_{ij})$ between the spins at $i$ -th and
$j$ -th sites.



It should be noted that due to the transverse field (the noncommuting
component of the cooperative part of the
Hamiltonian eqn. \ref{skd1}), the quantum dynamics of the spins
$\sigma^z$ arises from the Heisenberg
equation of motion. However, one can still
expect a simplified version of the dynamical
evolution in the mean-field approximation
(following refs. 
\cite{deGennes1963,Brout1966,bkc,SenChakrabarti1989,SuzukiKubo1968,AcharyyaPhysicaA1994,AcharyyaJPhysA1994}).

The mean-field Hamiltonian can be written as

\begin{equation}
 H(t)= - \sum_i  \vec{h}_i^{\text{eff}}(t)\cdot\vec{m_i}(t). 
 \label{skd2}
\end{equation}

\noindent Here, the effective mean-field
$\vec{h}_i^{\text{eff}}(t)$ on any spin
$\vec{m_i}$ 
has two
parts, the first part ${\vec{h}_i}^{z\,\text{eff}}(t)$
comes from the standard cooperative interaction (Curie-Weiss
type) among
the spins and the second part 
${{h}_i}^{x\,\text{eff}}(t)$ is the transverse field.

\begin{equation}
   {\vec{h}_i^{\text{eff}}}(t)= {{h}^{z\,\text{eff}}_i}(t)\hat{z}+{h_i}^{x\,\text{eff}}(t)\hat{x}
   \label{skd3}
\end{equation}

\noindent where
\begin{equation}
\left|\vec{h}_i^{\text{eff}}(t)\right| = \left[ \left( \sum_j J_{ij} m_j^{z}(t) +h(t) \right)^2 + \Gamma^{2}(t) \right]^{1/2} 
\label{skd4}
\end{equation}
\noindent with 
\begin{equation}
   h_i^{z\,\text{eff}}(t)= \sum_j J_{ij} m_j^{z}(t) + h(t) 
   \label{skd5}
\end{equation}

\noindent and 
\begin{equation}
h_i^{x\,\text{eff}}(t) = \Gamma(t).
\label{skd11}
\end{equation}

\noindent Here, $\vec{m}_i^{} \equiv \langle\vec{\sigma}_i^{}\rangle$, where $< \cdot >$
denotes the thermal average.


The generalized mean-field dynamics of the
Ising spins in the presence of both longitudinal
and transverse field, extending the classical
Suzuki–Kubo formalism \cite{SuzukiKubo1968} can be represented
(cf.  \cite{das2,DasPRE2025}) by the following
differential equation:

\begin{equation}
 \frac{d\vec{m}_i^{}}{dt} = - \vec{m}_i^{} +
\tanh\left(\frac{|\vec{h}_i^{\text{eff}}|}{T}\right) \frac{\vec{h}_i^{\text{eff}}}{\left|\vec{h}_i^{\text{eff}}\right|}.
\label{skd6}
\end{equation}

\noindent The above vector differential equation is basically
first order nonlinear coupled differential equations
for $\vec{m}_i^{} = \langle\vec{\sigma}_i^{} \rangle$.
They can be explicitly rewritten as


\begin{equation}
\frac{d m_i^{x}}{dt} = - m_i^{x} +
\tanh\left( \frac{ \left| \vec{h}_i^{\text{eff}} \right| }{T} \right) \cdot \frac{\Gamma} { \left| \vec{h}_i^{\text{eff}} \right| }
\label{skd7}
\end{equation}

and


\begin{equation}
\frac{d m_i^{z}}{dt} = - m_i^{z} +
\tanh\left( \frac{ \left| \vec{h}_i^{\text{eff}} \right| }{T} \right) \cdot
\frac{ h_i^{z\,\text{eff}}  }{ \left| \vec{h}_i^{\text{eff}} \right| }.
\label{skd8}
\end{equation}

For discrete time ($t$), the above
equation can be simplified to:


\begin{equation}
m_i^{x}(t+1) =
\tanh\left(
\frac{ \left| \vec{h}_i^{\text{eff}}(t) \right| }{T(t)}
\right)
\cdot
\frac{ \Gamma(t) }{ \left| \vec{h}_i^{\text{eff}}(t) \right| }
\label{skd9}
\end{equation}

and

\begin{equation}
m_i^{z}(t+1) =
\tanh\left(
\frac{ \left| \vec{h}_i^{\text{eff}} (t)\right| }{T(t)}
\right)
\cdot
\frac{ h_i^{z\,\text{eff}} (t) }{ \left| \vec{h}_i^{\text{eff}} (t)\right| },
\label{skd10}
\end{equation}

\noindent where $\left| \vec{h}_i^{\text{eff}}(t) \right|$, $h_i^{z\,\text{eff}}(t)$ are given by Eqs. (\ref{skd4}) and (\ref{skd5}) with the corresponding variables at time $t$. 



\subsection{Quantum Ising Hysteresis and Dynamic Transition}

\begin{figure}[h]
\begin{center}

\resizebox{8cm}{!}{\includegraphics[angle=0]{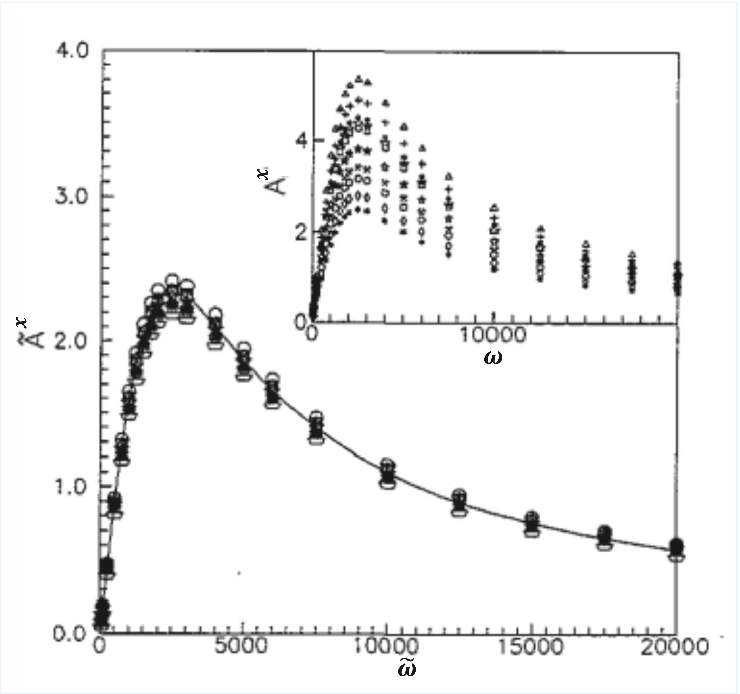}}
\caption{The variation of scaled loop area $\tilde{A}^x(\equiv A^x \Gamma_a^{-\alpha} T^{\beta})$ with the scaled frequency
$\tilde{\omega}\hspace{0.1cm}(\equiv{{\omega} \over {\Gamma_a^{\gamma} T^{\delta}}})$. The inset shows unscaled data ($A^x$) plotted against the frequency ($\omega$) for different values of $\Gamma_a$ and $T$. Adapted from \cite{AcharyyaJPhysA1994}}
\label{q-hyst}
\end{center}
\end{figure}

We have studied \cite{AcharyyaJPhysA1994} the hysteresis and dynamic phase transition in transverse Ising ferromagnet by solving the
mean field dynamical equation (for $h=0$ eqn. (\ref{skd6})). In this case the transverse field is sinusoidally oscillating in time
$\Gamma(t) = \Gamma_a {\rm cos}(\omega t)$, where $\omega = 2 \pi f$ and $f$ denotes the frequency of oscillation. The time dependence of $m^x(t)$ has been found to show a phase
difference with that of $\Gamma(t)$. This phase difference has given rise to the hysteretic response. The hysteresis
loop area $A^x = \oint m^x(t) d\Gamma$ has been studied as function of the frequency ($\omega$) and amplitude ($\Gamma_a$)
of the oscillating transverse magnetic field and the temperature ($T$) of the system. Here, for evaluating $A^x$
(and also for evaluating the dynamic order parameter $Q$), integration
over a complete cycle corresponds to the time period given by
$f^{-1}$. A  scaling form 
$A^x \sim \Gamma_a^{\alpha} T^{-\beta} g \left({{\omega} \over {\Gamma_a^{\gamma} T^{\delta}}}\right)$ 
has
been found through data collapse (Fig.~\ref{q-hyst}). The estimated (best fit)  values of the exponents are $\alpha=1.75\pm0.05$,
$\beta=0.50\pm0.02$, $\gamma=0.00\pm0.02$ and $\delta=0.00\pm0.02$. The scaling function $g(x) \sim {{x} \over 
{1+(cx)^2}}$.

\newpage

\begin{figure}[h]
\includegraphics[angle=0,width=7.5cm]{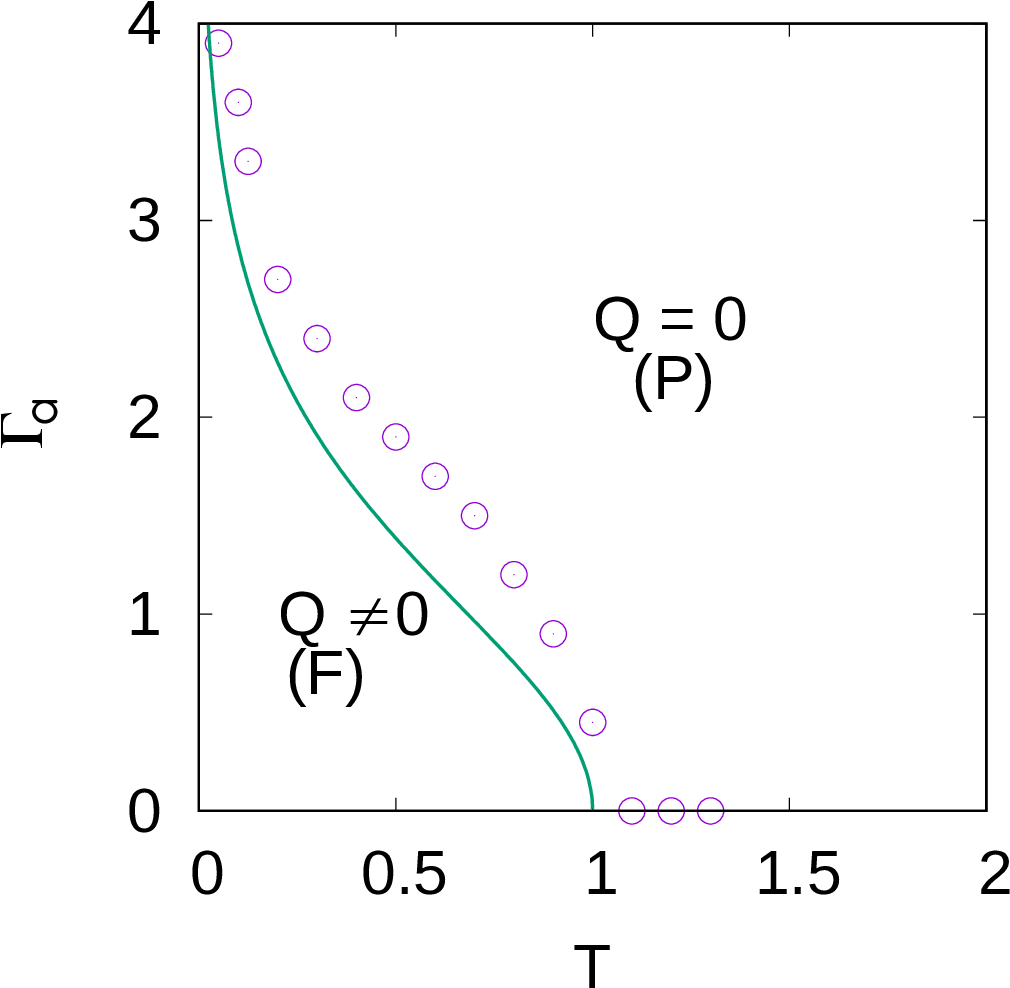}
\caption{The phase diagram for the dynamical phase transition: below the critical $\Gamma_a^c(T)$ line,
indicated by the symbols (o), the order parameter $Q$ acquires a nonzero value in the `F' phase and $Q=0$ in the
`P' phase above the critical line. Here, for the numerical data $\omega=2\pi\times500$. The continuous curve represents the approximate
phase boundary $T = {{\pi \Gamma_a/2} \over {{{\rm sinh}(\pi \Gamma_a/2)}}}$. Adopted from \cite{AcharyyaJPhysA1994}.}
\label{acharyya1994}
\end{figure}

The hysteretic response is found to be associated to another dynamical response, the dynamical phase transition. 
The time average magnetization (longitudinal component $m^z$) over the full cycle of the oscillating magnetic field, i.e., $Q = {{\omega} \over {2\pi}}\oint m^z(t) dt$,
transits to a nonzero value for specific set of values of the temperature ($T$) and the amplitude ($\Gamma_a$) of
the oscillating transverse magnetic field. A comprehensive phase boundary has been drawn and shown in Fig.~\ref{acharyya1994}. The approximate analytic form of the phase boundary has also been proposed to be
as $T = {{\pi \Gamma_a/2} \over {{{\rm sinh}(\pi \Gamma_a/2)}}}$. This analytic form of the dynamical phase boundary
matches well with that obtained numerically.


\subsection{ Quantum Ising Heat Engine}

\begin{figure}[h]
\begin{center}

\resizebox{7.5cm}{!}{\includegraphics[angle=0]{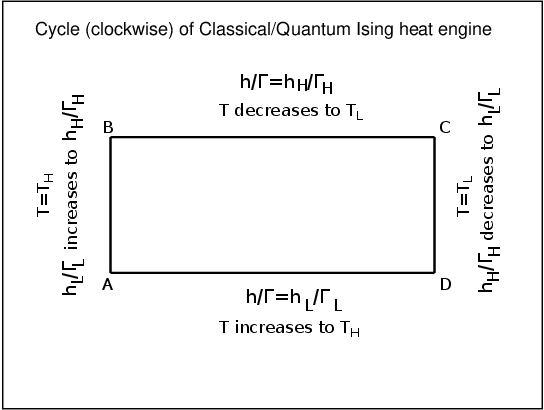}}
\caption{
Schematic diagram of a cycle of the heat engine: It starts from A
and returns to A after a clockwise rotation following the strokes AB, BC,
CD and DA on the working fluid (here an Ising system) of the engine. For
both classical and  quantum Ising heat engine, stroke AB corresponds to
fixed high temperature $T = T_H$ of the source and stroke CD corresponds
to fixed low temperature $T = T_L$ of the sink. In the stroke AB, the
transverse field $\Gamma$ changes from $\Gamma_L$ to $\Gamma_H$ for a
quantum Ising engine (while for the classical engine the longitudinal
field $h$ would change from $h_L$ to $h_H$). During the stroke BC the
temperature decreases from $T_H$ to $TL$ (for both kinds engines). During
the stroke  CD, the transverse field $\Gamma$ changes from $\Gamma_H$ to
$\Gamma_L$ for a quantum Ising engine (while the longitudinal field $h$
changes from $h_H$ to $h_L$ for the classical heat engine). In the fourth
stroke DA, the temperature of the working fluid changes from $T_L$ to
$T_H$ (for both classical and quantum Ising engines)}
\label{schematic}
\end{center}
\end{figure}

Following some of the original studies and results for increased
efficiencies in quantum heat engines, where the working fluid of the
engine is a quantum many-body system (see e.g., \cite{alicki,feldmann}), and some of the
extensive reviews (see e.g., \cite{mukherjee,bhatta}) on quantum heat engines with
standard quantum condensed matter working fluid, and the paper \cite{picc}
considering a transverse field Ising chain as the working fluid in a Otto
engine, a mean field quantum Ising system was employed as working fluid in
\cite{acharyya3}. Following \cite{acharyya3}, the mean field eqn. (\ref{skd6}) has been employed here to
compare the efficiencies of a four-stroke-cycle classical (when driven
along the heat cycle by the longitudinal field on the Ising system) and
quantum (when driven in time along the heat cycle by transverse field, in
general in presence of time  independent longitudinal field) heat engines,
designed in the following form (see Fig. \ref{schematic}):


\noindent (i) Stroke- A$\to$B: Field (transverse field
$\Gamma$, in quantum case) increases
linearly with time from a low value,
$\Gamma_L$, to a high value,
$\Gamma_H$) at a constant
high temperature $T_H$ of the heat
bath. Heat is being absorbed by the engine
during this stroke. This is an isothermal ($T_H$ constant) stroke. In this isothermal stroke, the system is 
allowed to absorb the heat
from the reservoir. The internal energy of the system (working fluid of the quantum engine) increases in this
stroke. 

\noindent (ii) Stroke-B$\to$C: The system is being cooled (linearly from $T_H$ to $T_L$) with thermalization (with the cold
bath or heat sink at temperature $T_L$) where the field remains fixed. 

\noindent (iii) Stroke-C$\to$D: This is another isothermal (at $T_L$) stroke where the field decreases
linearly from $\Gamma_H$ to $\Gamma_L$). The heat is being released hence the internal energy is found to decrease.

\noindent (iv) Stroke-D$\to$E: The temperature increases linearly (from $T_L$ to $T_H$) for fixed value of $\Gamma_L$.

Thus, in steady state, the system (working fluid, quantum Ising
system here) returns to the original or initial starting point of the cycle
(state) after the completion of the whole cycle. In the schematic diagram
(Fig. \ref{schematic}), the states denoted by A and E are the same with respect to all the
thermodynamic parameters.

The efficiency of the quantum engine would be clearly evaluated  from the
changes in internal energy (of the cooperative part of the Ising
Hamiltonian) absorbs heat during the steady state strokes A→B and that
released during the stroke C→D in Fig. 4.  Denoting the  heat absorbed and
released by $E_{absorbed}$ and $E_{released}$ respectively, we we can
express $E_{absorbed} = U(B) - U(A)$, where $U(A)$ and $U(B)$ represent
the magnitudes of the internal energy at the state points A and B
respectively. Similarly, $E_{released} = |U(D) - U(C)|$ (to keep is
positive), where $U(D)$ and $U(C)$ represent the magnitudes of the
internal energy at the state points D and C respectively. Hence, the
efficiency of the engine is $\eta = (E_{absorbed} - E_{released})/E_{absorbed}$.

Following \cite{acharyya3},  the numerical solutions (with parameter values used) and
the estimated values of internal energies $U$ of the quantum Ising system
(engine working fluid) at the points A, B, C and D, we get the efficiency
of the  quantum Ising heat engine, for different values of (fixed amount)
longitudinal field, as shown in Fig. \ref{quantum-efficiency}. The study could compare the
efficiencies of both classical and quantum Ising heat engines  (for
identical temperatures of the heat bath and heat sink). Fig. \ref{quantum-efficiency} shows that
the engine efficiency can be considerably enhanced (and brought closer to
the Carnot value corresponding to the temperatures of heat source and
sink) by putting a fixed optimal value for the longitudinal field. An
approximate
mean field type analysis supports \cite{acharyya3} such an observation.


\begin{figure}[h]
\begin{center}
\resizebox{7.5cm}{!}{\includegraphics[angle=0]{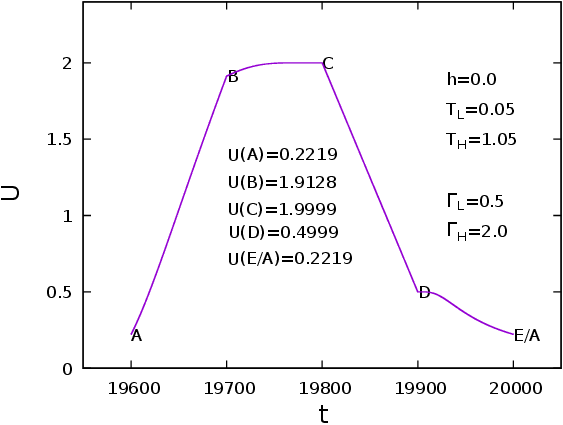}}
\caption{Numerically 
 estimated steady state internal energy ($U = {m^{z}}^2 + \Gamma m^x$) of the
working fluid (quantum Ising magnet), plotted against time ($t$) over a
full steady state cycle (for longitudinal field $h = 0$, and in the
presence of transverse field $\Gamma$). Adapted from \cite{acharyya3}.}
\label{quantum-energy}
\end{center}
\end{figure}

\begin{figure}[h]
\begin{center}
\resizebox{7.5cm}{!}{\includegraphics[angle=0]{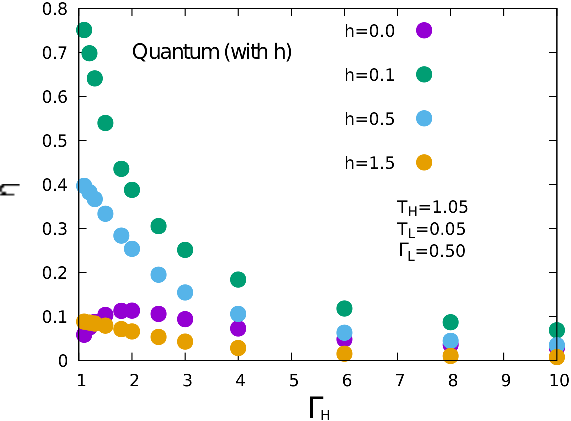}}
\caption{The efficiency $\eta$ of the quantum Ising
heat engine is plotted against the peak
value ($\Gamma_H$) of the transverse field
or quantum tunneling probability in the
Hamiltonian (\ref{skd1}) for different values (fixed over time)
of the longitudinal field ($h$). Note that $\Gamma_H = 1$ corresponds to the (equilibrium) quantum critical point (zero temperature) of the mean field system. Adapted from \cite{acharyya3}.}
\label{quantum-efficiency}
\end{center}
\end{figure}

\subsection{Quantum Annealing of the SK Model}


After the initial (1981) attempt \cite{bkc81} to study the  quantum Ising spin
glass of the Mattis and Edwards-Anderson variety in a transverse field,
the quantum Ising SK model was first studied in 1985 by Ishii and Yamamoto
\cite{ishii}, using high-temperature expansion of the mean field free energy from set
up by Thouless, Anderson and Palmer (TAP) type mean-field approximation
\cite{TAP1977}. We will discuss here the application of quantum annealing (see e.g.,
\cite{X1,das_rmp,tamura,rajak,ency}, \cite{kadwaki}), employing the discrete time version Eq.(\ref{skd9}) and Eq.(\ref{skd10}) of the
deGennes-Suzuki-Kubo equation and estimate the ground state energy of the
Sherrington-Kirkpatrick model \cite{SK1975,binder_young}. 
Here, we will put $h(t) = 0$ and
the interactions $J_{ij}$ in Hamiltonian Eq.(\ref{skd1}) are randomly ferromagnetic
and antiferromagnetic with zero average and have a Gaussian distribution:

\begin{equation}
P (J_{ij}) = (1/J)(N/2\pi)^{1/2} \exp[-(N/2)(J_{ij}/J)^2],         
\end{equation}

\noindent with 
\begin{equation}
[J_{ij}^2]_{av} - [J_{ij}]_{av}^2 = J^2/N = 1/N.
\end{equation}

\noindent Here, because of very weak average mean field on each spin, we
have to add a reaction term coming from the second order mean field
fluctuation (following TAP \cite{TAP1977} and the modification \cite{das2} required for the
low or zero temperature quantum spin-glass transitions here) in the
effective mean field component: In Eq.(\ref{skd5}) after incorporating the TAP reaction term $ h_i^{z\, \text{eff}}(t) =
\sum_j J_{ij}m_j^{z}(t) - \left[1 - q(t)\right]m_i^{z}(t)$ where $q=1/N\left[\sum_{i=1}^N\left(m^z_{i}\right)^2\right]_{av}$, while the
$x$-component in Eq.(\ref{skd11}) remaining the same, i.e.,  $h_i^{x\, \text{eff}}(t) =
\Gamma(t)$.
For quantum annealing,
the temperature and transverse field parameters will follow the annealing schedules:
\begin{equation}
    T(t) = T_0[1-t/\tau]
    \label{ann_temp}
 \end{equation} 
 
    \noindent and/or

\begin{equation}
    \Gamma(t) = \Gamma_0[1-t/\tau],  
    \label{ann_gamma}
\end{equation}

\noindent where at $t=0$, the temperature $T_0>T_c$ and/or $\Gamma_0>\Gamma_c$ and $T_c$, $\Gamma_c$ correspond to the temperature and transverse field values at the spin glass ($q\ne 0$) and para ($q=0$) phase boundary.  

We will discuss here (following \cite{das2}) the
numerical results obtained from the above-mentioned coupled equations.
We will solve numerically the above Eq. (\ref{skd10}) for given configurations of $m_i^{z}(t)$, using Eqs. (\ref{skd4}) and (\ref{skd5}) for $\vec{h}_i^{\text{eff}}(t)$, for an $N$ spin SK model ($N$ in the range 25 to 10,000) averaging over configurations in the range 10000 to 15, respectively.

For the classical case ($\Gamma$ = 0 with $h=0$), the equation
for $dm_i^z/dt$ has been already studied 
for classical annealing in the SK model \cite{DasPRE2025}.

\begin{figure}[h]
\includegraphics[width=8.5cm]{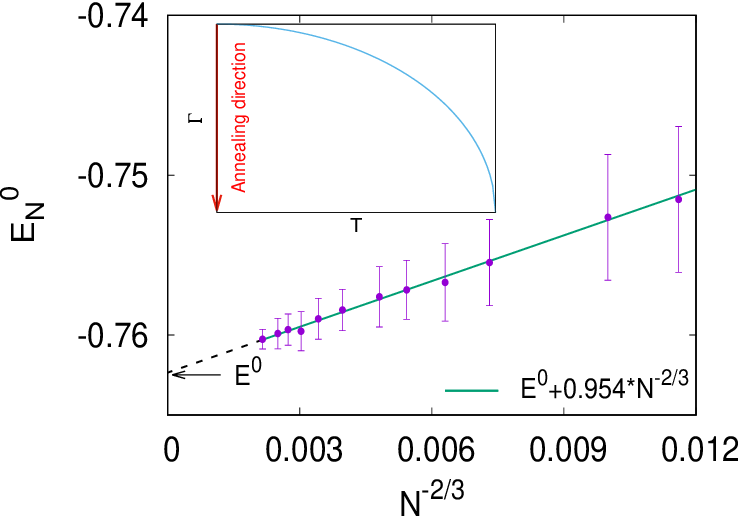}
\caption{Quantum annealing starting from the phase boundary point $\Gamma=\Gamma_0=1$ for $T=0$, $h=0$ with 
time variation of transverse field $\Gamma$ following Eq.(\ref{ann_gamma}): A schematic is shown in inset. The lowest energy values for given system size are plotted against $N^{-2/3}$ which shows a scaling $E^{0}_{N}\sim N^{-2/3}$. From the least-square fitting we get a ground state energy value (per spin) of $E^0=-0.7623\pm 0.0001$ (considering the exponent to be 2/3).
Adapted from \cite{das2}.}
\label{qm_en_q}
\end{figure}

\begin{figure}[h]
\includegraphics[width=8.5cm]{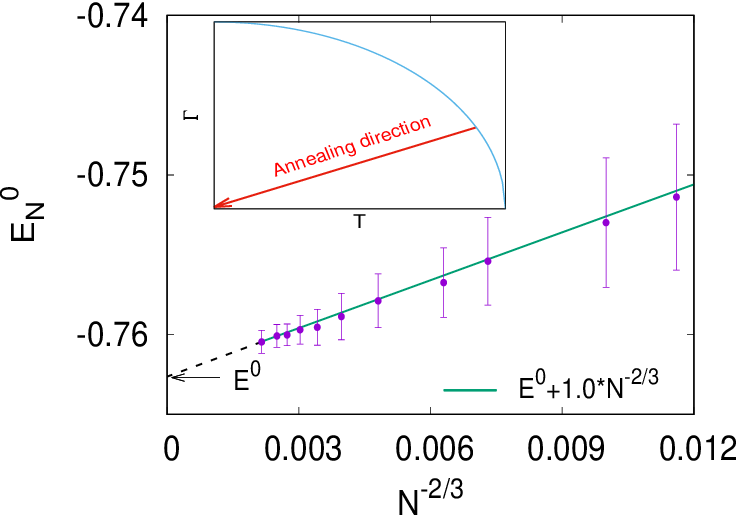}
\caption{Quantum Annealing starting from the critical phase boundary point $T=T_0=0.94$ and $\Gamma=\Gamma_0=0.5$ for $h=0$ where time variation of temperature and transverse field follow Eqs. (\ref{ann_temp}, \ref{ann_gamma}), a schematic is shown in inset: The lowest energy values for given system size are plotted against $N^{-2/3}$ that shows a scaling $E^{0}_{N}\sim N^{-2/3}$. From the least-square fitting we get a ground state energy value (per spin) of $E^0=-0.7626\pm 0.0001$ (considering the exponent to be $2/3$). Adapted from \cite{das2}.}
\label{qm_en_cq}
\end{figure}

The simplicity and affordability of the classical annealing using Suzuki-Kubo dynamics naturally raises interests in the quantum equivalent of the approach. For a pure quantum annealing (see Fig. \ref{qm_en_q}), the temperature is set equal to zero and the transverse field vanishes linearly with time, following Eq. (\ref{ann_gamma}). For a mixed annealing strategy (see Fig. \ref{qm_en_cq}), the initial values of the temperature $T_0$ and transverse field $\Gamma_0$ are set above $T_c=0.94$ and $\Gamma_c=0.5$ and they both vanish linearly with time following Eqs.(\ref{ann_temp}, \ref{ann_gamma}).

In Fig. \ref{qm_en_q}, the  ground state energy values obtained here for different system
sizes are very close to what was found for purely classical annealing \cite{boet,pala} for the corresponding sizes.  Indeed, the power law variations with
the system sizes also come out to be  the same for both the ground state
energies and their fluctuations. The linear annealing time $\tau$
used for the dynamics (in eqns. \ref{ann_temp} and  \ref{ann_gamma}) also has the same scaling
behavior as in the classical case. This implies that the algorithmic cost of
both  classical and quantum  simulations remain the same and scale with
$N^3$. The extrapolated value of the ground state energy per spin ($E^0=-0.7623\pm 0.0001$) in the infinite system size limit is also very close to the classical case ($E^0=-0.7629\pm 0.0002$).  Therefore, these results indicate that the algorithmic advantage here comes from the fact that the magnetization variables ($m_i$) are made continuous due to the Suzuki-Kubo dynamics. Hence, the case of quantum annealing, which effectively makes the magnetization variables continuous through both longitudinal and transverse fields, does not bring any added advantage.  


Finally, we look into the case of mixed annealing where both the temperature and transverse fields are varied simultaneously during the annealing process following Eqs. (\ref{ann_temp}) and (\ref{ann_gamma}). The time scale for annealing, $\tau$, is kept the same for both of these parameters. In Fig. \ref{qm_en_cq}, the ground state energy values and their fluctuations are plotted. The starting point of the annealing process was at $T=T_c= 0.94$ for $\Gamma=\Gamma_c=0.5$. Once again, the energy values are similar to what was obtained for the purely classical and quantum cases, with the same power law variations, albeit with a slightly different value for the constant $a$. The extrapolation to the thermodynamic limit gives a ground-state energy of $E^0=-0.7626\pm 0.0001$. Of course, the algorithmic cost also remains the same ($N^3$).

\section{Conclusions \& Discussion}

\begin{figure}[h]
\includegraphics[width=9.cm]{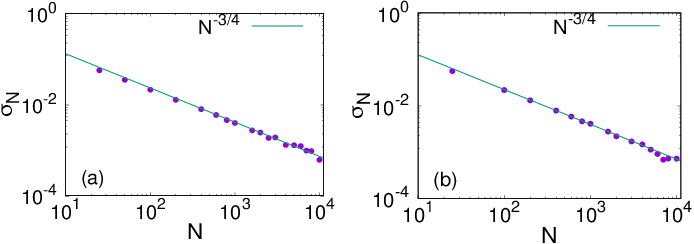}
\caption{The fluctuations $\sigma_N$ of $E^0_N$ $\bigg(\sigma_N\equiv\left[\sqrt{\left <{E^{0}_N}^2\right>-\left<E^{0}_N\right>^2}\right]_{av}\bigg)$ are plotted against system size. It shows a scaling $\sigma_N\sim N^{-3/4}$ for both (a) Quantum annealing and (b) mixed annealing.
Adapted from \cite{das2}.}
\label{qm_sd}
\end{figure}

Here, we have reviewed the recent development of a quantum many-body
theory of mean-field dynamics for cooperatively interacting Ising like
systems. For the static formulation, its origin could be traced back to
the pseudo-spin mapping of the BCS Hamiltonian and re-derivation  of the
BCS superconductivity theory in 1957 and its gap equations by Anderson (in
1958), where the mean-field self-consistent equations were employed for an XY (or
effectively Ising) model (with interaction strength given by those of the
Cooper-pairs)  in a transverse field (of strength given by the free electron
energy). That we have discussed in section II. Although the
quantum nature of the BCS superconductivity transition could be studied
from such a mapping, general nature of the phase transitions driven by
such transverse or quantum fields, was studied a little later, in
connection with the structural transitions in KDP type systems, starting
with de Gennes (1963). It was extended by Brout-Muller-Thomas (1966),
viewing the structural transitions as mode-softening of the collective
phonon modes coupled with the electron-tunneling modes in such systems,
incorporating the Suzuki-Kubo (1968) mean-field Ising dynamics in the
transverse Ising model (reviewed here in section III A). Such quantum phase transitions of the Ising models in transverse
fields were then extensively studied, also beyond mean field theories (see
e.g., \cite{bkc,inoue}, employing various analytical and computational
techniques from 1970 onwards. Some of these results are discussed here. First we have discussed the general formulation of quantum mean field dynamics, namely the deGennes-Suzuki-Kubo dynamics for transverse Ising like systems. Then, hysteretic response of an Ising system to a periodically varying
transverse or tunneling field, the scaling behaviour of the
hysteresis loop area
and the dynamic phase transition (separating the time averaged magnetization $Q$ from a non-zero value to zero; see Fig. \ref{acharyya1994}) has been discussed in section III B. In the following section (III C), the efficiencies of both classical and quantum
heat engines are compared using a quantum Ising model as working fluid and we find (see Fig. \ref{quantum-efficiency})
that the engine efficiency can be considerably enhanced (and brought
closer to the  Carnot value corresponding to the temperature of heat
source and sink) compared to the classical Ising system by putting a fixed
optimal value for the longitudinal field.  Finally we have discussed in
section III D, the  quantum annealing dynamics  for estimating the ground
state energy of the Sherrington-Kirpatrick spin glass model which gave a
computationally very fast algorithm,  leading to ground state energy comparable
with the best known analytic estimate. Here we have annealed the SK spin-glass system, first by changing the transverse field from 1 to 0 keeping the temperature at zero i.e., quantum annealing and later where both the transverse field and temperature are changed i.e., mixed annealing. We have gotten the ground state energy per spin (in the extrapolated limit) $E^0=-0.7623\pm 0.0001$ (see Fig. \ref{qm_en_q}) for quantum annealing and $E^0=-0.7626\pm 0.0001$ (see Fig. \ref{qm_en_cq}) for mixed annealing, with the fluctuation in $E^0$ reducing as $N^{-3/4}$ (see Fig. \ref{qm_sd}) with the system size $N$, compared to the best known analytic result $E^0=-0.763166 \dots$ \cite{Oppermann2007}. It
may be noted at this point that the mean field dynamics here for the
SK model, using both pure classical annealing (with tunneling field
$\Gamma$ set to zero) or pure quantum annealing (with
temperature $T$ set to zero), lead uniquely to an accurate
estimate ground state energy of the SK model .

Development of mean field theory for quantum many-body systems has a rich
history. It started with Nobel laureates Bardeen, Cooper and Schrieffer in
their 1957 theory of electron-phonon interaction induced superconductivity \cite{bardeen} and Nobel
laureate de Gennes' mean field theory (1963) of transverse Ising model \cite{deGennes1963} for proton
tunneling induced para-ferroelectric transition in KDP crystals. Nobel
laureate Anderson showed in 1958 \cite{anderson} that a pseudo-spin mapping of the the
BCS Hamiltonian leads to a cooperatively interacting XY
model (or effectively Ising model in Fermi vector space, near the Fermi
level) in a transverse or tunneling field (of strength given by the free
electorn dispersion). The resulting BCS gap parameter at zero temperature
($T = 0$; see Eqs. (\ref{appnew}, \ref{app4.9})) comes from the (pseudo-Ising) spin-spin
interaction. This non-vanishing gap $\Delta$ at $T < T_c$ for the electron energy
dispersion in the BCS superconductors would then induce,  according to the
phenomenological theory of superfluidity by Nobel laureate Landau (\cite{landau1},
1949), superconductivity  for the charged electrons.
Another Nobel Laureate
Muller (together with Brout and Thomas) \cite{Brout1966} in 1966 developed the mean
field or random phase approximation for the soft mode dispersion in the
transverse Ising model of KDP. The increase in transition temperature due
to reduced proton tunneling (increased mass) in deuterated KDP (see e.g.,
\cite{blc1,blc2}) could also be explained.  Following the initial study \cite{bkc81} of the
quantum Ising spin glass models (of the Mattis and Edwards-Anderson
variety), several attempts were made (see e.g., \cite{ishii,yokota}) to develop mean field theories for
the free energy of the transverse Ising SK model. These were mostly based
on the high temperature limit of the classical SK model TAP free energy
structure (proposed by Nobel laureate Thouless et al. in 1977 \cite{TAP1977}), which tried to incorporate the Replica Symmetry
Breaking effects in the ground state, as conceived by Nobel laureate
Parisi \cite{parisi79,parisi80} for the classical SK model. 

All these proved major advantage of the Mean field theory of transverse
Ising models in studying the static transition properties of long range
interacting quantum many body systems. Soon afterwards, the equilibrium
classical and quantum transition behavior in short-range interacting model
transverse field Ising systems started developing and became very well
established (see e.g., \cite{bkc,inoue,sachdev}). However, most of these developments
were for equilibrium or static phase transition behavior of the transverse
Ising systems. For applications of the model to non-equilibrium situations where the
fields on the condensed matter system  or even the temperature of the
system changes with time, for example in dynamic hysteresis in quantum
magnets (where the  transverse field on the quantum Ising system
changes periodically with time; see e.g., \cite{AcharyyaJPhysA1994,bkc-rev}), or in quantum
heat engines (with quantum Ising system as working fluid and both the
temperature and transverse field change with time during different
strokes of the engine cycles; see e.g., \cite{acharyya3}), or in quantum
annealing of the Sherrington-Kirkpatrick model (where the tunneling
field  and temperature both need to be annealed down to zero; see
e.g., \cite{das2}), one needs to augment the dynamics of the longitudinal and
transverse components (de Gennes \cite{deGennes1963})  of the Ising spins employing
the Suzuki-Kubo mean field dynamics \cite{SuzukiKubo1968}. This has been done in
section III A and the final equations for the longitudinal and transverse
spin components are given by Eqs.(\ref{skd7}) and (\ref{skd8}) in differential form and
by Eqs.(\ref{skd9}) and (\ref{skd10}) for discrete time evolution. Some of the important
results regarding the transverse magnetization loop area scaling and the
dynamic transition phase diagram (separating the  para or $Q = 0$ and
ordered ($Q > 0$, where $Q$ denotes the steady state value of the
integrated longitudinal magnetization over a complete period of the
transverse field), enhanced efficiency (approaching the Carnot value) of
the quantum Ising heat engine and the new quantum annealing algorithm for
estimating the ground state energy per spin of the SK model are discussed
in sections III B, III C and III D respectively. In particular, we note the
intriguing success of the mean field quantum Ising annealing algorithm
following the DeGennes-Suzuki-Kubo equation, in comfortably converging to
the best analytic estimate \cite{Oppermann2007} for the energy of the  Replica Symmetry
Broken \cite{parisi79,parisi80} ground state of the Sherrington-Kirkpatrick model \cite{SK1975}. It may be mentioned
here that the observations of similar accurate estimates for the ground
state energy as well of as the annealing times, as discussed in III D (see
\cite{das2,DasPRE2025} for details, see also \cite{stojnic}
for a recent comparision in the classical case), for both classical and quantum Ising
Sherrington-Kirkpatrick model suggest that “quantum supremacy” (cf. \cite{king})
gets merged in such cases when one employs the deGennes-Suzuki-Kubo mean
field dynamics for annealing. It may be
noted that the continuous nature of the magnetization variable (both
in classical and quantum cases) seems to be responsible here for
the major benefit in our annealing processes, where this dynamics
contributes equally for both the cases in the Sherrington-Kirkpatrick system
and did not show any "quantum supremacy". However,  it may further be
noted that in the quantum case the order parameter (magnetization) vector
has continuous dynamics for both the components  (longitudinal and transverse,
keeping the total magnetization magnitude constant)  and this feature (in
contrast to that in the classical case) can indeed help  more flexible search
(by navigating in higher dimension of the order parameter) for the ground state,
compared to the classical annealing in the context of different systems. Some
such interesting systems might be the cases of  sudden quench in BCS (see
e.g., \cite{capone,nava}), quantum quenching dynamics \cite{foster} of BCS superconductors to
Bose-Einstein (BE) condensate due to fast electromagnetic
perturbations and similar  many-body quantum dynamical systems. All these  indicate immense prospects for such mean field quantum many-body dynamical equation obtained by combining the de Gennes \cite{deGennes1963} mean field
quantum Ising model and the Suzuki-Kubo mean field (classical) Ising
dynamics as proposed here.

\acknowledgments{We are grateful to our long term collaborators on
these problems, namely  A. Dutta, J.-I. Inoue, P. Ray, P. Sen, R. B.
Stinchcombe, S. Suzuki and S. Tanaka. BKC is grateful to Indian National Science
Academy for a Senior Scientist Research Grant.}


\begin{thebibliography}{999}

\bibitem{bardeen}
J. Bardeen, L. N. Cooper and J. R.  Schrieffer, {\it Microscopic Theory of
Superconductivity},  Phys. Rev. {\bf 106}, pp. 162–164 (1957).
\url{https://doi.org/10.1103/PhysRev.106.162}

\bibitem{anderson}
 P. W. Anderson, {\it Random-Phase Approximation in the Theory of
Superconductivity}, Phys. Rev. {\bf 112}, pp. 1900-1916 (1958).
\url{https://doi.org/10.1103/PhysRev.112.1900}

\bibitem{deGennes1963}
 P. G. de Gennes, {\it Collective motions of hydrogen bonds}, Solid State
Commun. {\bf 1}, pp. 132–137 (1963).
\url{https://www.sciencedirect.com/science/article/abs/pii/0038109863902126}

\bibitem{Brout1966}
R. Brout, K. A. Muller, and H. Thomas, {\it Tunnelling and collective
excitations in a microscopic model of ferroelectricity}, Solid State
Commun. {\bf 4}, pp. 507–510 (1966).
\url{https://www.sciencedirect.com/science/article/abs/pii/0038109866904121}

\bibitem{eliot1}
R. J. Elliott, G. A. Gehring, A. P. Malozemoff, S. R. P. Smith, W. S.
Staude and R. N. Tyte, {\it Theory of co-operative Jahn-Teller distortions in
DyVO4 and TbVO4}, J. Phys. C: Solid State Phys., {\bf 4}, L179 (1971). 
\url{https://doi.org/10.1088/0022-3719/4/9/013}

\bibitem{eliot2}
R. J. Elliott, P. Pfeuty and C. Wood, {\it Ising Model with a Transverse
Field}, Phys. Rev. Lett. {\bf 25}, 443 (1970).
\url{https://doi.org/10.1103/PhysRevLett.25.443}

\bibitem{pfu}
P. Pfeuty, {\it The One-Dimensional Ising Model with a Transverse Field},
Ann. Phys. {\bf 57}, 79-90 (1970).
\url{https://doi.org/10.1016/0003-4916(70)90270-8}

\bibitem{stin}
R. B. Stinchcombe, {\it Ising model in a transverse field :I. Basic theory},
J. Phys. C : Solid State Phys., {\bf 6}, 2459-2483 (1973).
\url{http://iopscience.iop.org/0022-3719/6/15/009}


\bibitem{bkc}
B. K. Chakrabarti, A. Dutta and P. Sen, {\it Quantum Ising Phases and 
Transitions in Transverse Ising Models}, Springer, Heidelberg (1996).

\bibitem{barbar}
M. N. Barbar and P. M.  Duxbury, {\it A quantum Hamiltonian approach to
the two-dimensional axial next-nearest-neighbour Ising model}, J. Phys. A:
Math. Gen. {\bf 14}, L251DOI (1981).
\url{https://doi.org/10.1088/0305-4470/14/7/006}

\bibitem{SenChakrabarti1989}
P. Sen and B. K. Chakrabarti, {\it Ising models with competing axial
interactions in transverse fields}, Phys. Rev. B {\bf 40}, pp. 80–82 (1989).
\url{https://doi.org/10.1103/PhysRevB.40.760}

\bibitem{fisher}
D. S. Fisher, {\it Random transverse field Ising spin chains}, Phys. Rev.
Lett. {\bf 69}, 534 (1992).
\url{https://doi.org/10.1103/PhysRevLett.69.534}

\bibitem{sen}
P. Sen,  {\it Quantum phase transitions in the Ising model in a spatially
modulated field}, Phys. Rev. E {\bf 63}, 016112 (2000).
\url{https://doi.org/10.1103/PhysRevE.63.016112}

\bibitem{bkc81}
B. K. Chakrabarti, {\it Critical behavior of the Ising spin-glass models
in a transverse field}, Phys. Rev.  B {\bf 24}, 4062 (1981).
\url{https://doi.org/10.1103/PhysRevB.24.4062}

\bibitem{SuzukiKubo1968}
M. Suzuki and R. Kubo, {\it Dynamics of the Ising model near critical point
– I}, J. Phys. Soc. Jpn. {\bf 24}, pp. 51–60 (1968)
\url{https://doi.org/10.1143/JPSJ.24.51}

\bibitem{AcharyyaPhysicaA1994}
M. Acharyya and B. K. Chakrabarti, {\it Magnetic hysteresis loops as
Lissajous plots of relaxationally delayed response to periodic field
variation}, Physica A {\bf 202}, pp. 467–481 (1994)
\url{https://doi.org/10.1016/0378-4371(94)90473-1}

\bibitem{AcharyyaJPhysA1994}
M. Acharyya, B. K. Chakrabarti, and R. B. Stinchcombe, {\it Hysteresis in
Ising model in transverse field}, J. Phys. A: Math. Gen. {\bf 27}, pp.
1533–1540 (1994)
\url{https://doi.org/10.1088/0305-4470/27/5/018}

\bibitem{acharyya3}
M. Acharyya and B. Chakrabarti, {\it Quantum Ising heat engines: a mean
field study}, Eur. Phys. J. B  {\bf 97}, art. 45 (2024)
\url{https://doi.org/10.1140/epjb/s10051-024-00681-9}

\bibitem{X1}
P. Ray, B. K. Chakrabarti, A. Chakrabarti, {\it Sherrington-Kirkpatrick model in a transverse field: Absence of replica symmetry breaking due to quantum fluctuations},
Phys. Rev. B {\bf 39}, 11828-11833 (1989).
\url{https://doi.org/10.1103/PhysRevB.39.11828}

\bibitem{das_rmp}
A. Das and B. K. Chakrabarti, {\it Quantum Annealing and Analog Quantum
Computations}, Rev. Mod. Phys. {\bf 80}, 1061-1081 (2008)
\url{https://doi.org/10.1103/RevModPhys.80.1061}

\bibitem{tamura}
S. Tanaka, R. Tamura and B. K. Chakrabarti, {\it Quantum Spin
Glasses, Annealing and Computation}, Cambridge University
Press, Cambridge (2017).

\bibitem{rajak}
A.  Rajak, S. Suzuki, A. Dutta and B. K. Chakrabarti, {\it Quantum annealing: an overview}, Phil.
Trans. Royal  Soc. A  {\bf 381}, 20210417 (2023).
\url{https://doi.org/10.1098/rsta.2021.0417}

\bibitem{ency}
B. K. Chakrabarti and S. Mukherjee, {\it Quantum annealing and
computation}, Encyclopedia of Condensed Matter Physics (2nd Ed.)
{\bf 4},  536-542 (2024). \url{https://doi.org/10.1016/B978-0-323-90800-9.00057-3}


\bibitem{SK1975}
D. Sherrington and S. Kirkpatrick, {\it Solvable Model of a Spin-Glass},
Phys. Rev. Lett. {\bf 35}, pp. 1792-1796 (1975)
\url{https://doi.org/10.1103/PhysRevLett.35.1792}

\bibitem{binder_young}
K. Binder and A. P. Young, {\it Spin glasses: Experimental facts, theoretical concepts and open questions}, Rev. Mod. Phys. {\bf 58}, 801-976 (1986).
\url{https://doi.org/10.1103/RevModPhys.58.801}

\bibitem{Oppermann2007}
R. Oppermann, M. J. Schmidt, and D. Sherrington, {\it Double Criticality
of the Sherrington-Kirkpatrick Model at T = 0}, Phys. Rev. Lett. {\bf 98},
art. 127201 (2007)
\url{https://doi.org/10.1103/PhysRevLett.98.127201}

\bibitem{das2}
S. Das, S. Biswas and B. K. Chakrabarti, {\it Quantum Annealing in SK
Model Employing Suzuki-Kubo-deGennes Quantum Ising Mean Field Dynamics}, Eur. Phys. J. B {\bf 98}, 226 (2025).
\url{https://doi.org/10.1140/epjb/s10051-025-01073-3}

\bibitem{DasPRE2025}
S. Das, S. Biswas, and B. K. Chakrabarti, {\it Classical Annealing of
Sherrington-Kirkpatrick Spin Glass Using Kubo-Suzuki Mean-field Ising
Dynamics}, Phys. Rev. E {\bf 112}, art. 014104 (2025).
\url{https://doi.org/10.1103/www8-3ts1}



\bibitem{landau1}
L. D. Landau, {\it On the Theory of Superfluidity}, Phys. Rev.{\bf 75}, 884 (1949). 
\url{https://doi.org/10.1103/PhysRev.75.884}

\bibitem{landau2}
L. D. Landau and E. M. Lifshitz, {\it Course of Theoretical Physics}, Vol. 5,
Statistical Physics (1st ed.), Pergamon Press, Oxford (1951).


\bibitem{kittel}
C. Kittle, {\it Quantum theory of Solids}, John Wiley, New York (1987).


\bibitem{alicki}
 R. Alicki, {\it The quantum open system as a model of the heat engine}, J.
Phys. A Math. Gen. {\bf 12}, L103 (1979).
\url{https://doi.org/10.1088/0305-4470/12/5/007}

\bibitem{feldmann}
  T. Feldmann, R. Kosloff, {\it Quantum four-stroke heat engine:
thermodynamic observables in a model with intrinsic friction}, Phys. Rev. E
{\bf 68}, 016101 (2003).
\url{https://doi.org/10.1103/PhysRevE.68.016101}

\bibitem{mukherjee}
V. Mukherjee, U. Divakaran, {\it Many-body quantum thermal machines}, J.
Phys. Condens. Matter {\bf 33}, 454001 (2021).
\url{https://doi.org/10.1088/1361-648X/ac1b60}

\bibitem{bhatta}
S. Bhattacharjee, A. Dutta, {\it Quantum thermal machines and batterie}.
Eur. Phys. J. B {\bf 94}, 239 (2021).
\url{https://doi.org/10.1140/epjb/s10051-021-00235-3}

\bibitem{picc}
G. Piccitto, M. Campisi, D. Rossini, {\it The Ising critical quantum Otto
engine}, New J. Phys. {\bf 24}, 103023 (2023).
\url{https://doi.org/10.1088/1367-2630/ac963b}

\bibitem{ishii}
H Ishii and T Yamamoto, {\it Effect of a transverse field on the spin
glass freezing in the Sherrington-Kirkpatrick model}, J. Phys. C: Sol. St.
Phys. {\bf 18}, 6225 (1985).
\url{https://doi.org/10.1088/0022-3719/18/33/013}

\bibitem{TAP1977}
D. J. Thouless, P. W. Anderson, and R. G. Palmer, 
{\it Solution of ‘Solvable model of a spin glass’}, 
Phil. Mag. \textbf{35}, 593 (1977).
\url{https://doi.org/10.1080/14786437708235992}

\bibitem{kadwaki}
 T. Kadwaki and H. Nishimori,  {\it Quantum annealing in the transverse
Ising model}, Phys. Rev. E {\bf 58}, 5355 (1998).
\url{https://doi.org/10.1103/PhysRevE.58.5355}


\bibitem{boet}
S. Boettcher, {\it Extremal optimization for Sherrington-Kirkpatrick spin glasses}, Eur. Phys. J. B {\bf 46}, 501 (2005).
\url{https://doi.org/10.1140/epjb/e2005-00280-6}

\bibitem{pala}
M. Palassini, {\it Ground-state energy fluctuations in the Sherrington–Kirkpatrick model}, J. Stat. Mech.:Theor. Expt. {\bf P10005}, (2008).
\url{https://doi.org/10.1088/1742-5468/2008/10/P10005}

\bibitem{inoue}
S. Suzuki, J.-I. Inoue and B. K. Chakrabarti, {\it Quantum Ising Phases
and Transitions in Transverse Ising Models}, Springer-Verlag, Heidelberg
(2013).



\bibitem{blc1}
R. Blinc, {\it On the isotopic effects in the ferroelectric behaviour of
crystals with short hydrogen bonds}, J. of Phys. Chem. of Solids, {\bf 13}
204-211 (1960).
\url{https://doi.org/10.1016/0022-3697(60)90003-2}

\bibitem{blc2}
R. Blinc and B. Žekš, {\it Proton order-disorder in KH2PO4-type
ferroelectrics: Slater theory and ising model in a transverse tunneling
field}, Ferroelectrics, {\bf 72}, 193-227 (1987).
\url{http://dx.doi.org/10.1080/00150198708017947}

\bibitem{yokota}
T. Yokota, {\it Phase diagram for the infinite-range transverse Ising
spin-glass model}, Phys. Rev. B, {\bf 40}, 9321-9323 (1989).
\url{https://doi.org/10.1103/PhysRevB.40.9321}


\bibitem{parisi79}
G. Parisi, {\it Infinite Number of Order Parameters for Spin-Glasses},
Phys. Rev. Lett. 43, 1754 (1979).
\url{https://doi.org/10.1103/PhysRevLett.43.1754}

\bibitem{parisi80}
G. Parisi, {\it A sequence of approximated solutions to the S-K model for
spin glasses}, J. Phys. A 13, 1101 (1980).
\url{https://doi.org/10.1088/0305-4470/13/4/009}

\bibitem{sachdev}
S. Sachdev, {\it Quantum Phase Transitions (2nd Ed.)}, Cambridge University
Press, Cambridge (2011).

\bibitem{bkc-rev}
B. K. Chakrabarti and M. Acharya, {\it Dynamic transitions and hysteresis},
Rev. Mod. Phys. {\bf 71} , 847-849  (1999).
\url{https://doi.org/10.1103/RevModPhys.71.847}

\bibitem{stojnic}
M. Stojnic, {\it Binary perceptron computational gap -- a parametric
fl RDT view}, arXiv preprint.
\url{https://doi.org/10.48550/arXiv.2511.01037}

\bibitem{king}
A. D. King, et al., {\it Beyond-classical computation in quantum
simulation}, Science, 388, 199-204 (2025).
\url{https://www.science.org/doi/full/10.1126/science.ado6285}

\bibitem{capone}
 F. Peronaci, Marco Schiró, and M. Capone, {\it Transient Dynamics of
d-Wave Superconductors after a Sudden Excitation}, Phys. Rev. Lett.
{\bf 115}, 257001 (2015).
\url{https://doi.org/10.1103/PhysRevLett.115.257001}

\bibitem{nava}
 A. Nava, C. A. Perroni, R. Egger, L. Lepori, and D. Giuliano,
{\it Lindblad master equation approach to the dissipative quench dynamics
of planar superconductors},Phys. Rev. B {\bf 108}, 245129 (2023).
\url{https://doi.org/10.1103/PhysRevB.108.245129}

\bibitem{foster}
E. A. Yuzbashyan, M. Dzero, V. Gurarie, and M. S. Foster, {\it Quantum
quench phase diagrams of an s-wave BCS-BEC condensate}, Phys. Rev. A
{\bf 91}, 033628 (2015).
\url{https://doi.org/10.1103/PhysRevA.91.033628}


\end{thebibliography}
\end{document}